\documentclass[acmsmall, authorversion, anonymous=False]{acmart}

\usepackage{booktabs} 
\usepackage{subcaption}
\usepackage{cleveref}
\usepackage{xcolor}

\usepackage[ruled]{algorithm2e} 

\begin{document}
\title{Who Falls for Online Political Manipulation?}
 \subtitle{The case of the Russian Interference Campaign in the 2016 US Presidential Election}

\author{Adam Badawy}
\affiliation{%
  \institution{Information Sciences Institute, University of Southern California}}
\email{abadawy@usc.edu}

\author{Kristina Lerman}
\affiliation{%
  \institution{Information Sciences Institute, University of Southern California}}
\email{lerman@isi.edu}

\author{Emilio Ferrara}
\affiliation{%
 \institution{Information Sciences Institute, University of Southern California}}
\email{emiliofe@usc.edu}

\begin{abstract}
Social media, once hailed as a vehicle for democratization and the promotion of positive social change across the globe, are under attack for becoming a tool of political manipulation and  spread of disinformation. A case in point is the alleged use of trolls by Russia to spread malicious content in Western elections. This paper examines the Russian interference campaign in the 2016 US presidential election on Twitter. Our aim is twofold: first, we test whether predicting users who spread trolls' content is feasible in order to gain insight on how to contain their influence in the future; second, we identify features that are most predictive of users who either intentionally or unintentionally play a vital role in spreading this malicious content. We collected a dataset with over 43 million elections-related posts shared on Twitter between September 16 and November 9, 2016, by about 5.7 million users. This dataset includes accounts associated with the Russian trolls identified by the US Congress. Proposed models are able to very accurately identify users who spread the trolls' content (average AUC score of 96\%, using 10-fold validation). We show that political ideology, bot likelihood scores, and some activity-related account meta data are the most predictive features of whether a user spreads trolls' content or not.
\end{abstract}


\maketitle

\section{Introduction}
The initial optimism about the role of social media as a driver of social change has been fading away, following the rise in concerns about the negative consequences of malicious behavior online. Such negative outcomes have been particularly evident in the political domain. The spread of misinformation \citep{shorey2016,tucker2017} and the increasing role of bots \citep{bessi2016} in the 2016 US presidential elections has increased the interest in automatic detection and  prediction of malicious actor activity.

In this study, we focus on the role of Russian trolls in the recent US presidential elections. Trolls are usually described as users who intentionally ``annoy'' or ``bother'' others in order to elicit an emotional response. They post inflammatory messages to spread discord and cause emotional reactions \citep{phillips2015}. In the context of the 2016 US election, we define trolls as \emph{users who exhibit a clear intent to deceive or create conflict}. Their actions are directed to harm the political process and cause distrust in the political system. Our definition captures the new phenomenon of paid political trolls who are employed by political actors for a specified goal. The most recent and important example of such phenomenon is the Russian ``troll farms''---trolls paid by the Russian government to influence conversations about political issues aimed at creating discord and hate among different groups  \citep{gerber2016}.

Survey data from the Pew Research Center \citep{gottfried2016} show that two-thirds of Americans get their news from Social Media.  Moreover, they are being exposed to more political content written by ordinary people than ever before. \citet{bakshy2015} report that 13\% of posts by Facebook users---who report their political ideology---are political news. This raises the question of how much influence the Russian trolls had on the national political conversation prior to the 2016 US election, and how much influence such trolls will have in the upcoming elections. Although we do not discuss the effect that these trolls had on the political conservation prior to the election, we focus our efforts in this paper on the following two questions:  

\begin{description}
\item{RQ1:} Can we predict which users will become susceptible to the manipulation campaign by spreading content promoted by Russian trolls?
\item{RQ2:} What features distinguish users who spread trolls' messages?
\end{description}

The goal of these questions is, first, to test whether it is possible to identify the users who will be vulnerable to manipulation and participate in spreading the messages trolls post. We refer to such users as \emph{spreaders} in this paper. Our second goal is to better understand  what distinguishes spreaders form non-spreaders. If we can predict who will become a spreader, we can design a counter-campaign, which might stop the manipulation before it achieves its goal.   

For this study, we collected Twitter data over a  period of seven weeks in the months leading up to the election. By continuously pulling the Twitter Search API for relevant, election-related content using hashtag- and keyword-based queries, we obtained a  dataset of over 43 million tweets generated by about 5.7 million distinct users between September 16 and  November 9, 2016. First, we cross-referenced the list of Russian trolls published by the US Congress with our dataset and found that 221 Russian trolls accounts exist in our dataset. Next, we identified the list of users who retweeted the trolls. We gather important features about the users and use a machine learning framework to address the questions posed earlier. 

We used different machine learning classifiers on different models (each model includes a subset of the features, with the full model including all the features). We are able to achieve an average AUC score of 96\% for a 10-fold validation in terms of distinguishing spreaders form non-spreaders using Gradient Boosting for the full model on a subset of the dataset where the outcome variable has roughly equal number of spreaders vs. non-spreaders. Moreover, we verified our results on the full dataset as well as datasets where we increase the features but drop any rows with missing values. We are able to still achieve over 90\% average AUC score in the full model using Gradient Boosting. In terms of feature importance, political ideology is the most prominent for the balanced dataset, as well as in the validation settings. Number of followers, statuses (no. tweets), and bot scores were also in the top most predictive features both in the balanced and other datasets.

\section{Related Literature}
The use of trolls and bots in political manipulation campaigns around the globe is well documented through an array of reports by mainstream media outlets and academics (see \citet{tucker2018social} for a comprehensive review on the role of misinformation, bots, and trolls on social media). This phenomenon is not entirely new:  researchers warned about the potential for online  political manipulation for over a decade \citep{howard2006new,hwang2012socialbots}. Reports tracking and studying this phenomenon date back to the early 2010s \citep{ratkiewicz2011detecting,ratkiewicz2011truthy,metaxas2012social}. Since then, an increasing account of such events has been recorded in the context of several elections, both in the United States~\citep{bessi2016,kollanyi2016bots,shorey2016,woolley2016automation,woolley2016automating,marwick2017media,wang2016deciphering} and all over the world, including in South America \citep{forelle2015political,suarez2016influence}, the U.K. \citep{howard2016bots}, and Italy \citep{cresci2017paradigm}.

Although trolls do not necessarily need to be automated accounts, in many cases bots play a substantial role in political manipulation. \citet{bessi2016} report that 400k bots were responsible for posting 3.8 million tweets in the last month of the 2016 US presidential election, which is one-fifth of the total volume of online conversations they collected. Specifically, Russian political manipulation campaigns did not only target the US~\cite{badawy2018analyzing}: there is evidence of Russian interference in German electoral campaigns \citep{Applebaum2017make}, British elections \citep{gorodnichenko2018social}, and the Catalonian referendum~\citep{stella2018bots}. Russian-affiliated accounts were also reported in the 2017 French presidential elections, where bots were detected during the so-called \textit{MacronLeaks} disinformation campaign \citep{ferrara2017disinformation}. Moreover, a recent NATO report claims that around 70\% of accounts tweeting in Russian and directed at Baltic countries and Poland are bots. 

Russian political manipulation online did not stop at Russia's borders. Domestically, there is strong evidence that trolls and bots were present at multiple occasions. \citet{Ananyev2017} provides evidence of Russian government-affiliated trolls being able to change the direction of conversations on the LiveJournal blog, a popular platform in Russia in the 2000s. Moreover, the same entity that controlled many of the trolls studied in this paper, the Russian "troll factory", run by the \textit{Internet Research Agency}, had its trolls contribute to Wikipedia in support of  positions and historical narratives put forward by the current Russian government \citep{Labzina2017}. 

Online political manipulation is not only a Russian phenomenon. There is strong evidence of similar efforts by various governments to control political discussion online, particularly with the use of bots. \citep{king2017chinese} shows that the so-called "50-centers"--low-paid government workers who work online on behalf of the Chinese government-- try to distract Chinese citizens online form politically controversial topics. Even further, \citep{Miller2018} estimate that Chinese astroturfers produce about 15\% of all comments made on the 19 popular Chinese news websites. In Korea, bots were utilized as a part of a secret intelligence operation in support of the incumbent party's candidate reelection \citep{keller2017manipulate}.

{\color{black} There is strong evidence that political manipulation campaigns are on the rise, but how effective are they? In the case of the recent US presidential elections, \citet{allcott2017social} find that, even though ``fake news'' stories were widely shared on social media during the 2016 election, an average American saw only few of these stories. Despite this finding, we should not underestimate the potential role misinformation might play in distorting the views of citizens. Misinformed individuals hold consistently different opinions from those who are exposed to more factual knowledge. In some experimental studies, people who were exposed to accurate information about political issues often changed their views accordingly\citep{gilens2001political,sides2016electoral}. Other studies show that ignorance distorts collective opinion from what it would be if people were provided with more information about politics \citep{bartels1996uninformed,althaus1998information,gilens2001political} .

This distortion of individual opinion can lead to distortions at the aggregate level, as in the collective public opinion. On many occasions, these distortions might be initiated, encouraged, and exploited by domestic political elites or foreign powers. Manipulating actors are attempting to construct their own ``truth'' and push their version of the story in the public sphere to be adopted by their target audience. For example, some politicians might resort to distortion to win elections or avoid accountability for their performance in office \citep{fritz2004all,flynn2017nature}. Examples of misperceptions, whether caused by misinformation or not, that distort modern public policy debates in the US are abound. For example, US citizens hold drastically exaggerated perceptions about the amounts of the U.S. federal welfare, foreign aid, and the number of immigrants in the country. A recent Kaiser Family Foundation poll found that, on average, Americans estimated that 31\% of the federal budget goes to foreign aid, with very few people aware that the actual percentage does not exceed 1\% \citep{DiJulio2016}. Similarly, another survey found that fewer than one in ten respondents knew that welfare spending amounts to less than 1\% of the federal budget \citep{kuklinski2000misinformation}. Moreover, it has been found that Americans tend to overestimate the size of the immigrant population \citep{hopkins2018muted}. All these political misperceptions play a negative role in American political life and strengthen the already polarized environment that the US finds itself in right now. In case of the recent presidential elections, we could see that trolls were spreading misinformation about immigration, minority issues, and the government in general. The similarities with the above-mentioned cases of misinformation are obvious: trolls adopt classical techniques aimed at spreading misperceptions to push the agenda of the initiator of such campaigns. 

In recent years, growing ideological and affective polarization was accompanied by the increase in conspiracy theories and partisan misinformation. Belief in false and unsupported claims is frequently skewed by partisanship and ideology, suggesting that our vulnerability to them is increased by directionally motivated reasoning. Directionally motivated reasoning is defined as the tendency to selectively accept or reject information depending on its consistency with our prior beliefs and attitudes \citep{kunda1990case,taber2006motivated}.  This tendency makes the recent US presidential elections a good target for misinformation by internal and external actors: in an environment of severe polarization, both at the elite and mass level, false or misrepresented information that reinforces a person preexisting motivated perception/opinion can be quite effective. \citet{nyhan2010corrections} and \citet{flynn2017nature} show  that motivated reasoning can even undermine the effectiveness of corrective information, which sometimes fails to reduce misperceptions among vulnerable groups.}

\section{Data Collection}
\subsection{Twitter Dataset}

We created a list of hashtags and keywords that relate to the 2016 U.S. Presidential election. The list was crafted to contain a roughly equal number of hashtags and keywords associated with each major Presidential candidate: we selected 23 terms, including five terms referring to the Republican Party nominee Donald J. Trump (\#donaldtrump, \#trump2016, \#neverhillary, \#trumppence16, \#trump), four terms for Democratic Party nominee Hillary Clinton (\#hillaryclinton, \#imwithher, \#nevertrump, \#hillary), and several terms related to debates. To make sure our query list was comprehensive, we also added a few keywords for the two third-party candidates, including the Libertarian Party nominee Gary Johnson (one term), and Green Party nominee Jill Stein (two terms). 

By querying the Twitter Search API continuously and without interruptions between September 15 and November 9, 2016, we collected a large dataset containing 43.7 million unique tweets posted by nearly 5.7 million distinct users. Table \ref{Table 1} reports some aggregate statistics of the dataset. The data collection infrastructure ran inside an Amazon Web Services (AWS) instance to ensure resilience and scalability. We chose to use the Twitter Search API to make sure that we obtained all tweets that contain the search terms of interest posted during the data collection period, rather than a sample of unfiltered tweets. This precaution we took avoids known issues related to collecting sampled data using the Twitter Stream API that had been reported in the literature \cite{morstatter2013sample}.

\begin{table}[h]
\centering
\caption{Twitter Data Descriptive Statistics.}
\label{Table 1}
\begin{tabular}{ll}
Statistic                         & Count      \\ \midrule
\# of Tweets                     & 43,705,293 \\
\# of Retweets                   & 31,191,653 \\
\# of Distinct Users              & 5,746,997  \\ 
\# of Tweets/Retweets with a URL & 22,647,507
\end{tabular}
\end{table}

\subsection{Russian Trolls}

We used a list of 2,752 Twitter accounts identified as Russian trolls that was compiled and released by the U.S. Congress.\footnote{See https://www.recode.net/2017/11/2/16598312/russia-twitter-trump-twitter-deactivated-handle-list}  Table \ref{Table 2} offers some descriptive statistics of the Russian troll accounts. 
Out of the accounts appearing on the list, 221  exist in our dataset, and 85 of them produced original tweets (861 tweets). Russian trolls in our dataset retweeted 2,354 other distinct users 6,457 times. Trolls retweeted each other only 51 times. 
Twitter users can choose to report their location in their profile. Most of the self-reported locations of accounts associated with Russian trolls were within the U.S. (however, a few provided Russian locations in their profile), and most of the tweets were from users whose location was  self-reported as Tennessee and Texas (49,277 and 26,489 respectively). 
Russian trolls were retweeted 83,719 times, but most of these retweets were for three troll accounts only: \lq{TEN\_GOP}\rq, received 49,286 retweets; \lq{Pamela\_Moore13}\rq, 16,532; and \lq{TheFoundingSon}\rq, 8,755. These three accounts make up for over 89\% of the times Russian trolls were retweeted. Overall, Russian trolls were retweeted by 40,224 distinct users.

\begin{table}[h]
\centering
\caption{Descriptive Statistics on Russian trolls.}
\label{Table 2}
\begin{tabular}{@{}ll@{}}
\toprule
  & Value                                    \\ \midrule
\# of Russian trolls               & 2,735  \\
\# of trolls in our data           & 221    \\
\# of trolls wrote original tweets & 85     \\
\# of original trolls' tweets              & 861    \\
\bottomrule
\end{tabular}
\end{table}

\subsection{Spreaders}
Users who rebroadcast content produced by Russian trolls, hereafter referred to as \textit{spreaders}, may tell a fascinating story, thus will be the subject of our further investigation. Out of the forty thousand total spreaders, 28,274 of them produced original tweets (the rest only generated retweets). Overall, these twenty-eight thousand spreaders produced over 1.5 Million original tweets and over 12 Million other tweets  and retweets---not counting the ones from Russian trolls (cf., Table \ref{Table 3}). 

\begin{table}[h]
\centering
\caption{Descriptive statistics of spreaders, i.e., users who retweeted Russian trolls.}
\label{Table 3}
\begin{tabular}{@{}ll@{}}
\toprule
  & Value                                                          \\ \midrule
\# of spreaders                      & 40,224                   \\
\# of times retweeted trolls    & 83,719                   \\
\# of spreaders with original tweets & 28,274                   \\
\# of original tweets                & \textgreater 1.5 Million \\
\# of other tweets and retweets   & \textgreater 12 Million  \\ \bottomrule
\end{tabular}
\end{table}

\section{Data Analysis \& Methods}
In order to answer the questions posed in this paper, we gather a set of features about the users to \textit{(i)} predict the spreaders with the highest accuracy possible and \textit{(ii)} identify feature(s) which best distinguish spreaders from the rest. Table \ref{Table 4} shows all the features we evaluated in this paper, grouped under the following categories: Metadata, Linguistic Inquiry and Word Count (LIWC), Engagement, Activity, and Other variables.

\begin{table}[]
\centering
\caption{List of features employed to characterize users in our dataset}
\label{Table 4}
\begin{tabular}{@{}lllll@{}}
\toprule
Metadata         & LIWC             & Engagement        & Activity         & Other              \\ \midrule
\# of followers  & Word Count       & Retweet variables & \# of characters & Political Ideology \\
\# of favourites & Postive Emotion  & Mention variables & \# of hashtags   & Bot Score          \\
\# of friends    & Negative Emotion & Reply variables   & \# of mentions   & Tweet Count        \\
Status count     & Anxiety          & Quote variables   & \# of urls           \\
Listed count     & Anger            &                   &                  &                    \\
Default Profile  & Sadness          &                   &                  &                    \\
Geo-enabled      & Analytic         &                   &                  &                    \\
Background-image & Clout            &                   &                  &                    \\
Verified         & Affection        &                   &                  &                    \\
Account Age & Tone             &                   &                  &                    \\ \bottomrule
\end{tabular}
\end{table}

\begin{figure}

\includegraphics[width=1\textwidth]{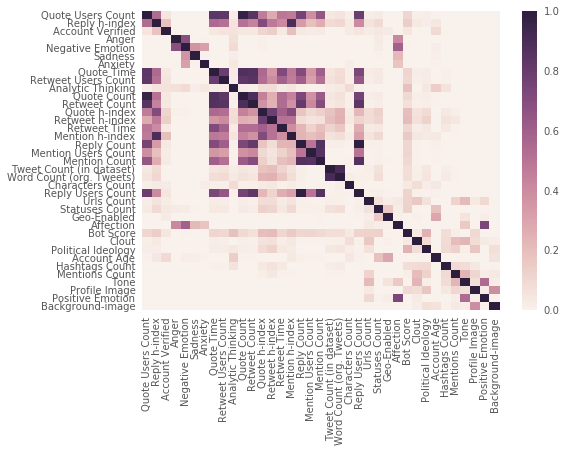}

  \caption{ Feature correlation heat map for the all users in the dataset.}
  \label{fig1}
\end{figure}

To understand what each variable in the Metadata and LIWC categories means, see the Twitter documentation page\footnote{ https://developer.twitter.com/en/docs/tweets/data-dictionary/overview/user-object} and \citep{pennebaker2015development}, respectively. The Activity variables convey the number of characters, hashtags, mentions, and URLs produced by users, normalized by the number of tweets they post. Tweet Count, under Other, is the number of user's tweets appearing in our dataset. The remaining variables are more involved and warrant a detailed explanation: we explain how Political Ideology, Bot Scores, and Engagement variables were computed in the following sections. One may wonder how much the features evaluated here correlate with each other, and whether they provide informative signals in terms of predictive power about the spreaders. Figure \ref{fig1} shows that, besides Engagement variables, most of the features are not highly correlated among each other (Pearson correlation is shown, results do not vary significantly for Spearman correlation). There are however a few notable exceptions: Word Count and Tweet Count, LIWC Positive Emotion and Affection, Anxiety and Anger---these pairs all show very high correlation. This is not surprising, considering that these constructs are conceptually close one another. As for the Engagement variables, we can see a "rich get richer" effect here, where users who have higher scores in terms of some of the sub-features in the Engagement category, are also higher in other sub-features. For example, by construction the \textit{Retweet h-index} will be proportional to the number of times a user is retweeted, and similarly for replies, quotes and mentions---all these Engagement features are explained in great detail in a section \S\ref{sec:engagement}.


\subsection{Political Ideology}

\subsubsection{Classification of Media Outlets}

We classify users by their ideology based on the political leaning of the media outlets they share. We use lists of partisan media outlets compiled by third-party organizations, such as AllSides\footnote{\url{https://www.allsides.com/media-bias/media-bias-ratings}} and Media Bias/Fact Check.\footnote{\url{https://mediabiasfactcheck.com/}} The combined list includes 249 liberal outlets and 212 conservative outlets. After cross-referencing with  domains obtained in our  Twitter dataset, we identified 190 liberal and 167 conservative outlets. We picked five media outlets from each partisan category that appeared most frequently in our Twitter dataset and compiled a list of users who tweeted from these outlets. The list of media outlets/domain names for each partisan category is reported in Table \ref{Table 5}.

\begin{table}[h]
\centering
\caption{Liberal \& Conservative Domain Names.}
\label{Table 5}
\begin{tabular}{@{}ll@{}}
\toprule
Liberal                & Conservative             \\ \midrule
www.huffingtonpost.com & www.breitbart.com        \\
thinkprogress.org      & www.thegatewaypundit.com \\
www.politicususa.com   & www.lifezette.com        \\
shareblue.com          & www.therebel.media       \\
www.dailykos.com       & theblacksphere.net       \\ \bottomrule
\end{tabular}
\end{table}

We used a polarity rule to label Twitter users as liberal or conservative depending on the number of tweets they produced with links to liberal or conservative sources. In other words, if a user had more tweets with links to liberal sources, he/she would be labeled as liberal and vice versa. Although the overwhelming majority of users include links that are either liberal or conservative, we remove any users that had equal number of tweets from each side\footnote{We use five categories, as in left, left center, center, right center, right, to make sure we have a final list of users who are unequivocally liberal or conservative and do not fall in the middle. The media outlet lists for the left/right center and center were compiled from the same sources.}---this to avoid the conundrum of breaking ties with some arbitrary rule. Our final set of labeled users include 29,832 users.

\subsubsection{Label Propagation}

We used \textit{label propagation}\footnote{We used the algorithm in the Python implementation of the IGraph library \cite{csardi2006igraph}} to classify Twitter accounts as liberal or conservative, similar to prior work~\cite{Conover2010predicting}. In a network-based label propagation algorithm, each node is assigned a label, which is updated iteratively based on the labels of the node's network neighbors. In  label propagation, a node takes the most frequent label of its neighbors as its own new label. The algorithm proceeds updating labels iteratively and stops when the labels no longer change (see \cite{raghavan2007} for more information). The algorithm takes as parameters \textit{(i)} weights, in-degree or how many times node $i$ retweeted node $j$;  \textit{(ii)} seeds (the list of labeled nodes). We fix the seeds' labels so they do not change in the process, since this seed list also serves as our ground truth. 

We construct a retweet network where each node corresponds to a Twitter account and a link exists between pairs of nodes when one of them retweets a message posted by the other. 
We use the 29K users mentioned in the media outlets sections as seeds, those who mainly retweet messages from either the liberal or the conservative media outlets in Table \ref{Table 5}, and label them accordingly. We then run label propagation to label the remaining nodes in the retweet network.


\begin{table}[h]
\centering
\caption{Breakdown for overall users, trolls and spreader by political ideology}
\label{Table 6}
\begin{tabular}{@{}lll@{}}
\toprule
             & Liberal            & Conservative     \\ \midrule
\# of users     & \textgreater 3.4 M & \textgreater 1 M \\
\# of trolls    & 107                & 108              \\
\# of spreaders & 1,991              & 38,233           \\ \bottomrule
\end{tabular}
\end{table}

To validate results of the label propagation algorithm, we applied stratified 5-fold cross validation to the set of 29K seeds. We train the algorithm on four-fifths of the seed list and test how it performs on the remaining one-fifth. The averge precision and recall scores are both over 91\%.

To further validate the labeling algorithm, we notice that a group of Twitter accounts  put media outlet URLs as their personal link/website. We compile a list of the hyper-partisan  Twitter users who have the domain names from Table \ref{Table 5} in their profiles and use the same approach explained in the previous paragraph (stratified 5-fold cross-validation). The average precision and recall scores for the test set for these users are above 93\%. Table \ref{Table 7} shows the average precision and recall scores for the two validation methods we use: both labeled  over 90\% of the test set users correctly, cementing our confidence in the performance of the labeling algorithm.

\begin{table}[h]
\centering
\caption{Precision \& Recall scores for the seed users and hyper-partisan users test sets.}
\label{Table 7}
\begin{tabular}{@{}lll@{}}
\toprule
 & Seed Users & Hyper-Partisan Users       \\ \midrule
Precision  & 91\%                 & 93\% \\
Recall     & 91\%                 & 93\% \\ \bottomrule
\end{tabular}
\end{table}

\subsection{Bot Detection} 

Determining whether either a human or a bot controls a social media account has proven a very challenging task \cite{ferrara2016rise, subrahmanian2016darpa}. We use an openly accessible solution called Botometer (a.k.a. BotOrNot) \cite{davis2016botornot}, consisting of both a public Web site (\url{https://botometer.iuni.iu.edu/})  and a Python API (\url{https://github.com/IUNetSci/botometer-python}), which allows for making this determination with high accuracy. Botometer is a machine-learning framework that extracts and analyses a set of over one thousand features, spanning six sub classes:
\begin{description}
\item [User]: Meta-data features that include the number of friends and followers, the number of tweets produced by the users,
profile description and settings.
\item [Friends]: Four types of links are considered here: retweeting, mentioning, being retweeted, and being mentioned. For each group separately, botometer extracts features about language use, local time, popularity, etc. 
\item [Network]: Botometer reconstructs three types of networks: retweet, mention, and hashtag co-occurrence networks. All networks are weighted according to the frequency of interactions or co-occurrences.
\item [Temporal]: Features related to user activity, including average rates of tweet production over various time periods and distributions of time intervals between events.
\item[Content]: Statistics about length and entropy of tweet text and Part-of-Speech (POS) tagging techniques, which identifies different types of natural language components, or POS tags. 
\item[Sentiment]: Features such as: arousal, valence and dominance scores \citep{warriner2013norms}, happiness score \citep{kloumann2012positivity}, polarization and strength \citep{wilson2005recognizing}, and emotion score \citep{agarwal2011sentiment}.
\end{description}

\begin{figure}

\includegraphics[width=.6\textwidth]{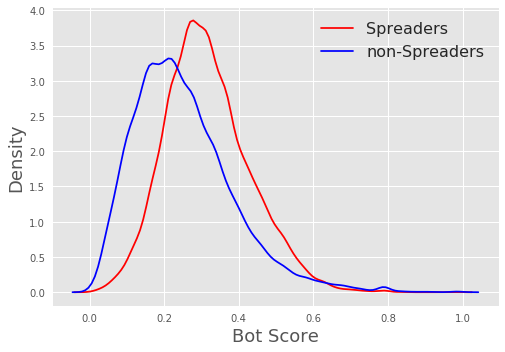}

  \caption{Probability density distributions of bot scores assigned to spreaders (red) and  non-spreaders (blue).}
  \label{fig2}
\end{figure}

We utilize Botometer to label all the spreaders, and we get bot scores for over 34K out of the total 40K spreaders. Since using Botometer to get scores all non-spreaders (i.e., over 5.7M users) would take an unfeasibly long time (due to Twitter's restrictions), we randomly sample the non-spreader user list and use Botometer to get scores for a roughly equivalent-size list of non-spreader users. The randomly-selected non-spreader list includes circa 37K users. To label accounts as bots, we use the fifty-percent threshold which has proven effective in prior studies \cite{davis2016botornot}: an account is considered to be a bot if the overall Botometer score is above 0.5. Figure \ref{fig2} shows the probability distribution for spreaders vs. non-spreaders. While most of the density is under the 0.5 threshold, the mean of spreaders (0.3) is higher than the mean of non-spreaders. Additionally, we used a t-test to verify that the difference is significant at the 0.001 level (p-value).

\begin{figure}
\begin{subfigure}{0.48\textwidth}
\includegraphics[width=\linewidth]{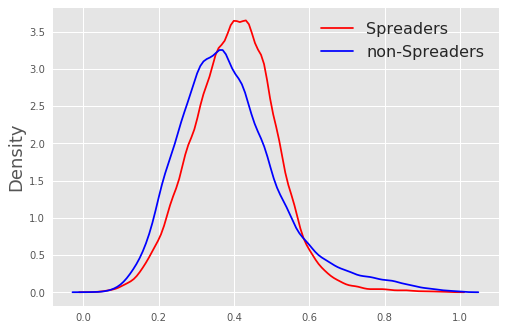}
\caption{Content}
\label{fig:3a}
\end{subfigure}\hspace*{\fill}
\begin{subfigure}{0.48\textwidth}
\includegraphics[width=\linewidth]{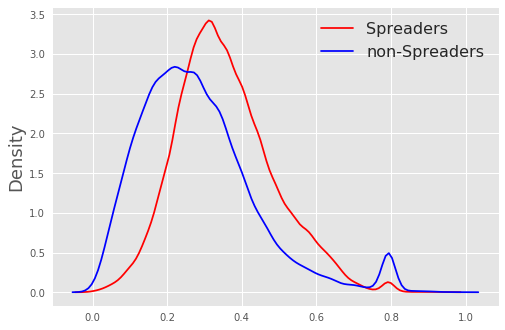}
\caption{Friend} 
\label{fig:3b}
\end{subfigure}

\medskip
\begin{subfigure}{0.48\textwidth}
\includegraphics[width=\linewidth]{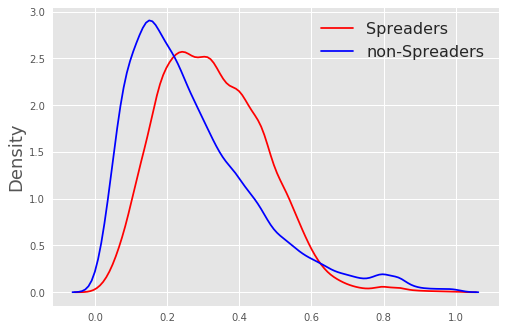}
\caption{Network}
\label{fig:3c}
\end{subfigure}\hspace*{\fill}
\begin{subfigure}{0.48\textwidth}
\includegraphics[width=\linewidth]{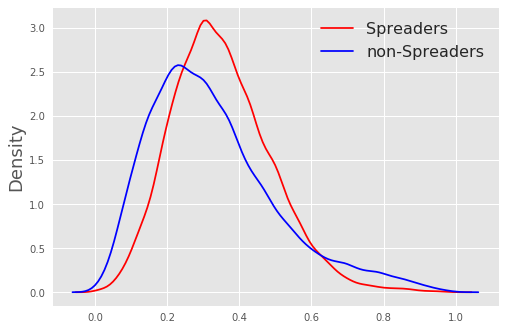}
\caption{Sentiment}
\label{fig:3d}
\end{subfigure}

\medskip
\begin{subfigure}{0.48\textwidth}
\includegraphics[width=\linewidth]{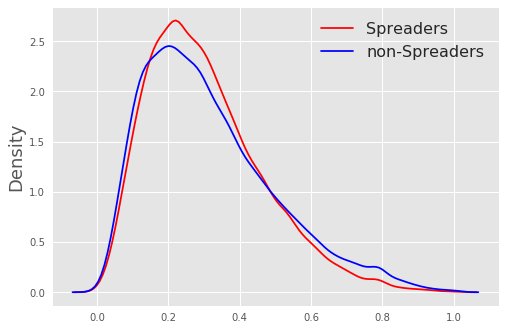}
\caption{Temporal} 
\label{fig:3e}
\end{subfigure}\hspace*{\fill}
\begin{subfigure}{0.48\textwidth}
\includegraphics[width=\linewidth]{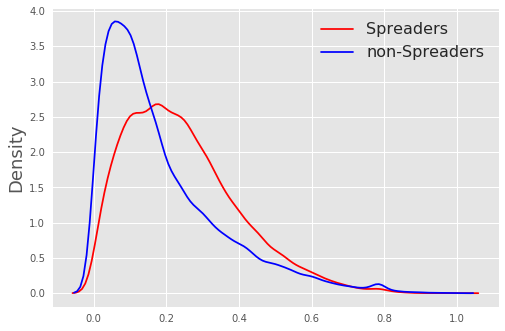}
\caption{User} 
\label{fig:3f}
\end{subfigure}

\caption{Distribution of the probability density of Botometer sub\-feature scores for spreaders vs. non\-spreaders.} \label{fig:3}
\end{figure}

As for the plots in Figure \ref{fig:3}, it is evident that the spreaders are different on almost all the Botometer subclass scores, except for the temporal features. The differences in all plots are statistically significant (p <0.001). Besides, looking at the distributions, we can see that the difference in user characteristics (metadata), friends, and network distributions, are substantively different as well. Moreover, the mean of spreaders is higher in all the subclass features.

\subsection{Engagement}\label{sec:engagement}
We plan to measure user engagement in four activities: retweets, mentions, replies, and quotes. 
Engagement of a user is measured through three components: the quantity, longevity, and stability in each activity. For instance, for a set of $N$  users, this measure would calculate the engagement index score of user $i\in N$ by including the following:

\begin{description}
  \item[1)] number of retweets, replies, mentions, and quotes by $N -i $ users for user $i$;

  \item[2)] time difference between the last and the first quote, reply, and retweet per tweet;

  \item[3)] consistency of mentioning, replying,  retweeting, and quoting by $N-i$ users for user $i$ across time (per day);

  \item[4)] number of unique users who retweeted, commented, mentioned, and quoted user $i$ 
  
\end{description}

Item three is  measured using h-index \citep{hirsch2005}. The measure captures two notions: how highly referenced and how continuously highly referenced a user is by other members in the network \citep{lietz2014}. This measure was originally proposed to quantify an individual's scientific research output. In this context, a user has index $h$ if for $h$ days, he/she is referenced at least $h$ times and in all but $h$ days no more than $h$ times. 

\section{Results}
Predicting spreaders on the original dataset may be considered a daunting task: only a relatively small fraction of users engaged with Russian trolls' content (about 40K out of 5.7M users). However, for the same reason, if a model were to trivially predict that no user will ever engage with Russian trolls, the model would be accurate most of the time (i.e., most users won't be spreaders), even if its recall would be zero (i.e., the model would never correctly predict any actual spreaders)---provided that we want to predict spreaders, this model would not be very useful in practice. In other words, our setting is a typical machine-learning example of a highly-unbalanced prediction task.

To initially simplify our prediction task, we created a balanced dataset that is limited to users who have bot scores.\footnote{Will get back to the original prediction task on the highly-unbalanced dataset  later in this section.}  
This balanced dataset has about ~72K users, with ~34K spreaders and ~38K non-spreaders. To test our ability to detect spreaders and to see which features are most important in distinguishing between the two groups, we leverage multiple classifiers and multiple models: the first model serves as a baseline with each model including more variables until we reach the full model. Since our goal was not that to devise new techniques, we used four off-the-shelf machine learning algorithms: Extra Trees, Random Forest, Adaptive Boosting, and Gradient Boosting.  We train our classifiers using Stratified 10-fold cross-validation with the following preprocessing steps \textit{(i)} replace all categorical missing values with the most frequent value in the column \textit{(ii)} replace missing values with the mean of the column. 

Table \ref{Table 8} shows all the models we evaluate, from the simplest baseline model (Metadata) to the full model that includes all the features we present in Table \ref{Table 4}.

\begin{table}[h]
\centering
\caption{Machine Learning Models from the Baseline (Metadata) to Full Model.}
\label{Table 8}
\begin{tabular}{@{}ll@{}}
\toprule
Model & Features                                         \\ \midrule
1     & Metadata                                        \\
2     & Metadata + LIWC                                 \\
3     & Metadata + LIWC + Activity                      \\
4     & Metadata + LIWC +Activity  + Engagement         \\
5     & Metadata + LIWC +Activity  + Engagement + Other \\ \bottomrule
\end{tabular}
\end{table}

\begin{figure}
\includegraphics[width=.8\textwidth]{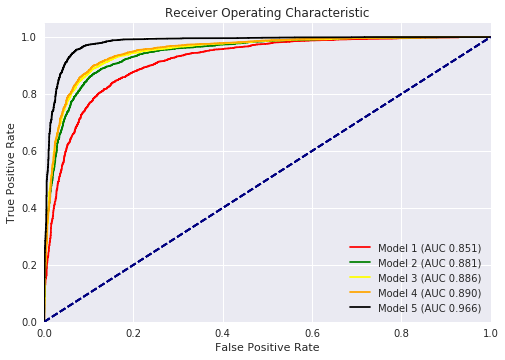}

  \caption{Area under the ROC curve plot for the five models under evaluation using Gradient Boosting. We show five models, in each we use the fold/model that yields the highest AUC among the trained ones. It is evident that the addition of bot scores, political ideology, and tweet count variables are important in improving the performance of the classifiers. The legend shows the average AUC scores for each model.}
  \label{fig4}
\end{figure}

\begin{figure}
\includegraphics[width=1\textwidth]{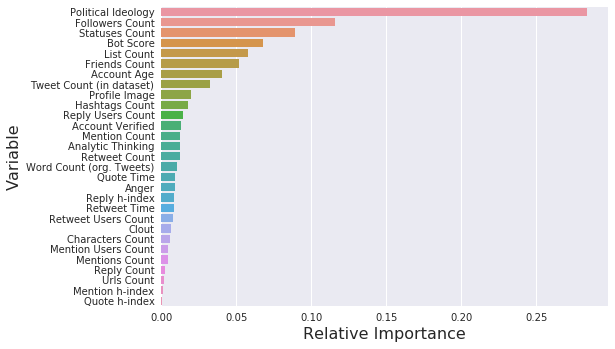}

  \caption{Relative importance of the features using Gradient Boosting for the full model (best performing fold) in predicting users as spreaders vs. non-spreaders. Political Ideology explains over 25\% of the variance, followed by Followers Count, Statuses Counts, and Bot Scores, each explaining roughly 5\% to 10\% of the variance.}
  \label{fig5}
\end{figure}

For Gradient Boosting, which is the best performing classifier among the four we evaluate, we obtained average AUC scores for the 10 folds that range from 85\% to 96\%. Figure \ref{fig4} shows the ROC curve plots for each model (using the fold/model with the highest AUC score among the trained ones). The jump from 89\% to 96\% for the AUC scores from Model 4 to 5 shows that the addition of bot scores and  political ideology are meaningful in distinguishing spreaders from non-spreaders (the legend in Figure \ref{fig4} shows the average AUC score for each model). To better understand the contribution of the features in predicting the target values (i.e., spreader vs. non-spreader), we look at the variable importance plot of the Gradient Boosting results for  Model 5. The \textit{Variable Importance} plot (cf., Figure \ref{fig5}) provides a list of the most significant variables in descending order by a mean decrease in the Gini criterion. The top variables contribute more to the model than the bottom ones and can discriminate better between spreaders and non-spreaders. In other words, features are ranked based on their predictive power according to the given model. Figure \ref{fig5} shows that, according to Model 5 and Gradient Boosting, political ideology is the most predictive feature, followed by number of followers, statuses/tweets count (obtained from the metadata), and bot score, in a descending order of importance. The plot does not show all the features, since the omitted features contribute very little to the overall predictive power of the model. 

\begin{figure}
\begin{subfigure}{0.50\textwidth}
\includegraphics[width=\linewidth]{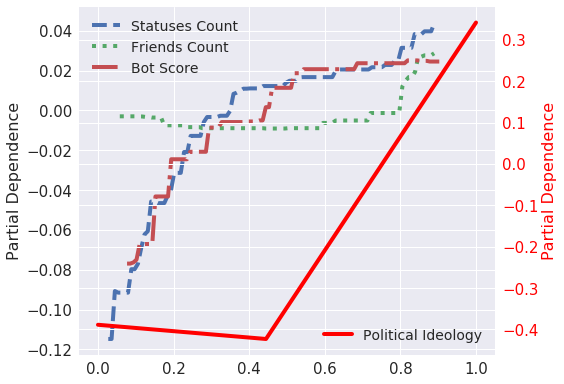}
\caption{Upward Trends}
\label{fig:6a}
\end{subfigure}\hspace*{\fill}
\begin{subfigure}{0.50\textwidth}
\includegraphics[width=\linewidth]{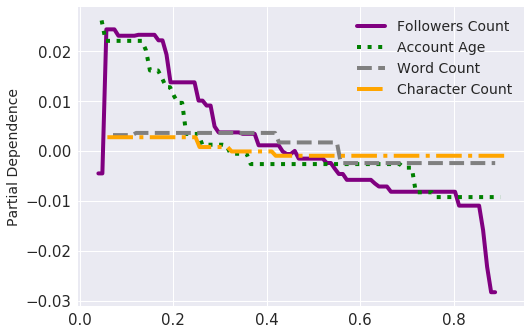}
\caption{Downward Trends} 
\label{fig:6b}
\end{subfigure}
\caption{Partial Dependence plots for some of the features considered in the full model (best preforming fold). These partial dependence plots are for the Gradient Boosting Classifier fitted to the balanced dataset. Each plot shows the dependence of feature  (spreader) on the feature under consideration, marginalizing over the values of all other features (Note: x-axis values are CDF-normalized).} 
\label{fig:6}
\end{figure}

Feature importance plots reveal which features contribute most to classification performance, but they do not tell us the nature of the relationship between the outcome variable and the predictors. Although predictive models are sometime used as black boxes, \textit{Partial Dependence} plots (cf., Fig. \ref{fig:6})
can tell us a lot about the structure and direction of the relationship between the target and independent variables. They show these relationships after the model is fitted, while marginalizing over the values of all other features. 
The dependency along the x-axis captures the range of a given feature, with that feature values normalized between 0 and 1.\footnote{Political ideology should be considered in the range from 0 (to identify left leaning users), to 1 (for right-leaning ones).}

Using Partial Dependence, we illustrate that the target variable (spreader) has positive relationships with the following features: political ideology, statuses count, bot scores, and friends count. 
Figure \ref{fig:6a} visualizes these relationships (we put political ideology on a different y-axis in order to show that its magnitude of influence on the target variable is significantly higher compared to all other features, including downward trend features in Figure \ref{fig:6b}). This suggests that moving from left to right political leaning increases the probability of being a spreader; larger number of posts, more friends (a.k.a. followees), and higher bot scores are also associated with higher likelihood of being a spreader.

On the other hand, we can see that the outcome variable has a negative relationship with followers count, account age, characters count, and word count, as shown in Figure \ref{fig:6b}. This means that having fewer followers, having a recently-created account, posting shorter tweets with fewer words, are all characteristics associated with higher probability of being a spreader.

Going back to the original highly-unbalanced dataset, we aim to validate the results above using two strategies: \textit{(i)} we run Gradient Boosting (with the same preprocessing steps) on the whole dataset of ~5.7M for the five models we outlined in table \ref{Table 8}; \textit{(ii)} we run Gradient Boosting classifier on models without imputations and with all missing observations deleted. For the first approach, the average AUC scores ranged form 83\% for the baseline model to 98\% for the full model. For the second approach, due to the sparsity of some features, the overall number of observations decreases significantly when these features are added. Putting the overall number of observations aside, the average ROC scores for a 10-fold validation for the roughly same set of models specified earlier range from 84\% to 91\%. In terms of feature importance, political ideology is again the most important in the full model, with status count and bot scores following it in importance.  In summary, the results above remain consistent when validating on the highly-unbalanced prediction task.

\section{Discussion and Limitations}
The results in previous section show that \textit{(i)} with some insight on users who spread or produce malicious content, we are able to predict the users that will spread their message to a broader audience; \textit{(ii)} in the case we focus on, the 2016 US presidential elections, political ideology was highly predictive of who is going to spread trolls' messages vs. not. Moreover, looking at the top predictive features in Figure \ref{fig5}, basic metadata features give a strong signal in terms of differentiating spreaders from non-spreaders, along with the bot score. Looking at the subclass features from Botometer, Figures \ref{fig:3b}, \ref{fig:3c}, and \ref{fig:3f} show that spreaders and non-spreaders are significantly different on the dimensions of friends, network, and user metadata, with spreaders having higher bot scores on all three (thus, they have a higher likelihood of being a bot according to those subclasses). 

Looking at the partial dependence plots, we can deduce that spreaders write a lot of tweets (counting retweets as well), have higher bot scores, and tend to be more conservative (conservative is labeled as the highest numerical value in the political ideology feature). Also, since the range of the y-axis tells us about the range of influence a feature has on the target value, it is evident that political ideology  has by far the most influence on distinguishing between spreaders and non-spreaders. On the other hand, we can also deduce that spreaders do not write much original content, tend not have that many followers, and  have more recently established user accounts. In the downward trends in Figure \ref{fig:6b}, we can see that followers count and account age have more influence on the target value in comparison to the other features in this plot.  

{\color{black} Although our analysis shows that certain features might be predictive of spreaders, there are certain limitations to how generalizable these findings can be. First and foremost, our data does not capture all the trolls in the trolls' list. The rest of the trolls might not be present in our dataset due to various reasons; perhaps, they were simply not that active during the period of our data collection. In any case, it is virtually impossible to gauge the effect of such missingness on our results and conclusions. Second, we lack sufficient information on how the troll list was compiled in the first place. This might be an issue, since the methodology taken to identify these trolls could include certain biases that might affect our conclusions. Third, while political ideology emerged as the most predictive feature among the ones included in the paper, it is important to note that a large portion of the trolls' tweets were targeting conservatives in the first place. Thus, this finding gives us some insight into the conservatives' reaction to political manipulation; however, it does not tell us much about the respective reaction on the liberal side. Fourth, certain tools used in the paper might work better on some types of data than others \citep{Hoffman:2017:ECA:3171581.3134687}. For example, some of LIWC categories might work better on longer and more elaborate texts than tweets, particularly on types of text that capture more sophisticated emotions beyond simply negative and positive valence. Lastly, and this goes for any case study, our conclusions could be influenced by the special circumstances of the 2016 US presidential elections, and the same phenomenon may or may not unfold in the similar manner in a different context. 

Despite the above-mentioned limitations of our approach and data, it is important to note that understanding massive online political manipulation campaigns is, nevertheless, extremely important, and that the threat of such attacks on our democratic systems will not go away any time soon. In this paper, we used a dataset collected through keywords that, in our view, fully encompass the political event under study. Using the list of trolls published by the Congress, we were able to study a phenomenon we do not yet fully understand, to comprehend how it functions, and to start a conversation on whether it can be stopped or prevented in a duly manner. Overall, we employed a variety of rigorous computational tools to analyze and predict trolls' activities in the recent US presidential elections.}

\section{Conclusion}
This work focused on predicting spreaders who fall for online manipulation campaigns. We believe that identifying likely victims of political manipulation campaigns is the first step in containing the spread of malicious content. Access to reliable and trustworthy information is a cornerstone of any democratic society. Declining trust of citizens of democratic societies in mainstream news and their increased exposure to content produced by ill-intended  sources poses a great danger to democratic life.  Social science literature shows a lot of evidence that mis-perceptions on the individual level can aggregate into a distortion in the collective  public opinion \citep{bartels2002beyond,baum2009shot}, which can have severe policy implications \citep{fritz2004all,flynn2017nature}. Thus, we believe that studying how and who spreads political manipulation content is extremely important, and it is an issue that many social media platforms should attempt to contain.

\begin{acks}
The authors gratefully acknowledge support by the Air Force Office of Scientific Research (award \#FA9550-17-1-0327). The views and conclusions contained herein are those of the authors and should not be interpreted as necessarily representing the official policies or endorsements, either expressed or implied, of AFOSR or the U.S. Government.
\end{acks}

\bibliographystyle{ACM-Reference-Format}
\bibliography{citations.bib}

\end{document}